\documentclass{ws-mpla}
\usepackage{url}

\begin{document}

\markboth{J. Gu\'ena, M. Lintz and M.-A. Bouchiat}{Atomic Parity Violation}

%
%

\title{ATOMIC PARITY VIOLATION :\\ PRINCIPLES, RECENT RESULTS, PRESENT MOTIVATIONS}

\author{\footnotesize J. GU\'ENA, M. LINTZ and M.-A. BOUCHIAT}

\address{  Laboratoire Kastler Brossel and F\'ed\'eration de Recherche, \\D\'epartement de Physique de
l'Ecole Normale Sup\'erieure,\\ 24 Rue Lhomond, F-75231  Paris  Cedex 05, France \\
guena@lkb.ens.fr}

\maketitle


\begin{abstract}
We review the progress made in the determination
of the weak charge, $Q_W$, of the cesium nucleus which raises the status of Atomic Parity Violation
measurements to that of a precision electroweak test. Not only is it necessary to
have a precision measurement of the electroweak asymmetry in the highly forbidden 6S-7S
transition, but one also needs a precise calibration procedure. 
The 1999 precision measurement by the Boulder group implied a 2.5 $\sigma$ deviation of $Q_W$ from the theoretical prediction. This triggered many particle physicist suggestions as well as examination by atomic theoretical physicists of several sources of corrections. After about three years the disagreement was removed without appealing to "New Physics".
 Concurrently, an original experimental approach was developed in our group for more than a decade. It is based on detection by stimulated emission with amplification of the left-right asymmetry. We present our decisive, recent progress together with our latest results. We emphasize the important impact for electroweak theory, of future measurements in cesium possibly pushed to the 0.1\% level. Other possible approaches are currently explored in several atoms. 

\keywords{Electroweak interference; nuclear weak charge; atomic parity violation; stimulated-emission detection.}
\end{abstract}

\ccode{PACS Nos.: 32.80.Ys, 11.30.Er, 42.50.Gy}

\section{Introduction}
\noindent In this short review paper we intend to 
survey the theoretical and experimental progress made in Atomic Parity Violation
since the emergence of the field in 1974. First ($\S\; 2$), we indicate the connection between Atomic Parity Violation (APV) and
Electroweak (EW) Theory. Everything centres around one electroweak parameter, the weak charge of an
atomic nucleus, $Q_W$, associated with the exchange of the $Z_0$ boson between the atomic electron and nucleus.
We provide a short background to APV experiments. 
How can $Z_0$ bosons affect  atomic radiative transitions? 
 How is $Q_W$ extracted from experiment? Besides the
measurement of a left-right asymmetry itself, $A_{LR}$, the theoretical interpretation of the result requires
state of the art atomic physics calculations. 
Then ($\S\; 3$) we present past experiments in atomic cesium, the paradigmatic atom for a quantitative test of the EW theory, and discuss their implications. As will be seen in $\S\; 4$, the apparent 
deviation between the most precise result by the Boulder group\cite{ben99,woo97} and the Standard Model (SM) prediction triggered important theoretical
efforts. It also gave additional motivations for our current Cs experiment in Paris, presented next. 
We show that this experiment now provides valuable PV data by a
method totally different from Boulder. We indicate other possible experimental approaches, in different atoms, under current investigation. Finally ($\S\; 5$), we dicuss the relevance today of APV experiments.


\section{Electroweak parity violation in atoms}
\begin{figure}[b]
\centerline{\psfig{file=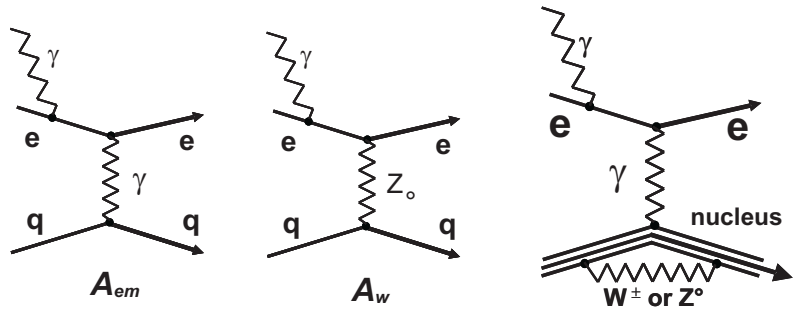,width=4.1in}}
\vspace*{3mm}
\caption{\small Left and center: Schematic representation of the two amplitudes, electromagnetic $A_{em}$ and weak $A_{W}$ associated with a photon or $Z_{0}$ exchange between electron e and quark q, which can contribute to the same
radiative process and give rise to an electroweak interference. Right: Schematic representation of the nuclear-spin-dependent PV interaction associated with
the nuclear anapole moment.}
\end{figure}
For a naive estimate of $A_{LR}$ let us consider two radiative electron-hadron processes (Fig. 1, Left and center), the first one of amplitude $A_{em}$ governed exclusively by electromagnetic interactions, and the second
one of amplitude $A_W$ associated with a $Z_0$ boson exchange. The weak
amplitude $A_W$ contains a part which is \textit{odd} under space reflexion, which will give rise to a left-right
asymmetry $A_{LR}$ by interference with the dominant electromagnetic amplitude.
For two
mirror-image experiments we obtain two different transition probabilities: $P_{L/R} = \vert
A_{em} \pm A_W^{odd} \vert ^2$ and we find:
\begin{equation}
 A_{LR}=\frac {P_L - P_R} {P_L + P_R}\simeq 2 \,Re\left(
 A_W^{odd} / A_{em} \right) \; .
\end{equation}
If $ q $ denotes the four-momentum transfer between the lepton and the
hadron, $ A_{ em} $ is proportional to $ e^2/q^2 $ while $ A_ {
W}\propto g^2/ \left(q^2+M^2_{Z_0} c^2 \right)$ with $g^2 \approx e^2$, hence $  A_{LR}  \simeq
q^2 / M^2_{Z_0} c^2$.  In atoms we expect $ q $ to be given by the inverse of the Bohr radius,
$\sim  m_e\alpha c$. We thus arrive at an exceedingly
small value of the asymmetry :
$$ A_{LR}\simeq \alpha^ 2{m^2_e \over M^2_{Z_0} }\approx 10^{-15} \; .$$
Such a result would appear to make the observation of the left-right
asymmetry in atoms completely hopeless. Fortunately there are important enhancement
mechanisms which make this naive estimate far too pessimistic. In fact, in actual experiments,
$A_{LR}$ can be as large as a few times $10^{-6}$.

The first source of enhancement finds its origin in the so-called $Z^3$ law. This law predicted
by C. Bouchiat and one of the authors (M.-A. B.) in 1974\cite{bou97,bou741} states that the electroweak
effects in atoms  should grow a little faster than the cube of the atomic number $ Z $.
Indeed, for valence electrons belonging to penetrating orbitals, like $s_{1/2}$ or $p_{1/2}$, the
orbital is deformed in the vicinity of the atomic nucleus, right where the short range
interaction takes place. It looks like the orbital associated with a Coulomb potential of
charge Z, whose radius is given by $a_0 / Z$. Hence, in heavy atoms the factor
$\vert  q^2 \vert $ is enhanced by the factor $Z^2$. Moreover, the various nucleons in the
atomic nucleus add their contributions coherently. Since, in heavy atoms, the number of
nucleons grows roughly as Z, the overall enhancement effect is proportional to $Z^3$.

For a more quantitative analysis it is necessary to introduce the parity violating
electron-nucleus potential which, in the non-relativistic limit, can be written as:
\begin{equation}
V_{pv}(r_{e})=\frac{Q_{W} \, G_{F}}{4\sqrt{2} }  \left( \,{\delta}^3(\vec{r_{e}}) \,
 {\vec{\sigma}}_{e} . {\vec{v}}_{e}/c \,
   +\; h.c.  \, \right) \; .
\end{equation}
Here we have kept the dominant contribution in which the $Z_0$ couples to the electron as
an axial vector and to the nucleons as a vector. Consequently, the strength of this interaction
is naturally expressed in terms of a nuclear charge, $Q_W$. For the electroweak interaction
this charge plays the same role as the nuclear electric charge for the Coulomb interaction.
Hence its name, the "weak nuclear charge". Like the electric nuclear charge, $ Z$, the weak
charge $ Q_{W}$
 is the sum of the weak charges of all the constituents of
the atomic nucleus, the $u$ and $d$ quarks\nobreak\ :
\begin{equation}
 Q_ { W}=(2Z+N)Q_ { W}(u)+(Z+2N)Q_ { W}(d) \; .
\end{equation}
In the Standard Model, with the electroweak parameter ${\sin}^2\theta_{W}\simeq 0.23$, it so happens that $Q_{W}$ lies close to the neutron number\footnote{\small This formula is valid only to lowest order
in the electroweak interaction.}\nobreak\ :
\begin{equation}
 Q_ { W}(SM)=-N-Z \left(4 \;{\sin}^2\theta_{W}-1 \right) \simeq -N \;
\end{equation}
 
The second source of enhancement comes from the possibility of exciting highly forbidden
transitions like the $6S_{1/2} \rightarrow 7S_{1/2}$ transition in cesium. In a
transition such as $6S_{1/2} \rightarrow 7S_{1/2}$ the electromagnetic selection rules strictly
forbid the existence of an electric dipole transition. The weak interaction associated with $Z_0$
exchange breaks this rule and gives rise to a parity violating electric dipole
amplitude\footnote{\small Note that $Z_0$ exchange gives rise to a transition electric dipole, not to a
static electric dipole. The reason is that this PV weak interaction preserves time reversal
invariance, while a static dipole would be a manifestation of simultaneous P and T violation.
Another consequence of time reversal invariance is that $E_1^{pv}$ is pure imaginary in a phase
convention where $M_1$ is real.\cite{bou97}},
$E_1^{pv}$. This $E_1$ amplitude is of course very small,
$\simeq 10^{-11}$ atomic units ($ea_0$, $a_0$ Bohr radius). On the
other hand, symmetry allows the existence of a magnetic dipole amplitude,
$M_1$. However, because the two states connected by the transition have different radial
quantum numbers, the $M_1$ transition is suppressed:
 its amplitude is only $4 \times 10^{-5} \; \mu_B/c$, so that we can anticipate a relatively large
asymmetry: $Im\; E_1^{pv}/M_1
\simeq 0.5 \times 10^{-4}$. Even today, cesium
appears to be a good compromise between a high atomic number necessary to have sizable
effects and the simple atomic structure required to make precise atomic calculations.

Once the $ Z^3 $ enhancement became apparent, the main question was
how best to take advantage of it.
There were in fact two different
lines of attack. The first takes advantage of highly
forbidden $ M_1 $ transitions such as $6S_{1/2} \rightarrow 7S_{1/2}$ in cesium. Its
merits are the relatively large asymmetry and the
simple atomic structure characteristic of an alkali atom which has
a single valence electron outside a tight atomic core. This type
of transition, however, represented completely new territory and the suppression factor
looked absolutely huge, $ \sim 10^{14} $, so that one could anticipate
difficulties with the signal-to-noise ratio. Nevertheless, this was the approach chosen by our
group in  Paris and later on by the Boulder group\cite{gil85}. The forbidden $ M_1 $ $
6P_{1/2}-7P_{1/2}
$ transition in Tl was selected by the Berkeley  group\cite{neu77}.
The second line of attack
consists in working with allowed $ M_1 $ transitions in atoms of even
higher $ Z $ such as Tl, Pb and Bi, with Z respectively equal to 81, 82, and 83. The suppression
factor is only $10^5$ and {\it a priori} this should avoid the signal-to-noise
difficulties. This approach was adopted at Oxford\cite{mac91}, Seattle\cite{mee93},
Novosibirsk\cite{bar80} and Moscow\cite{bir84}. Precise measurements in these elements have
been achieved, but presently the difficulty lies in the more complicated atomic structure. The
accuracy in
$Q_W$ is today limited by the precision in atomic calculations. For this reason this
presentation is limited to the case of cesium where the most precise determination has been so
far achieved.  Hereafter, we shall concentrate on the $6S_{1/2}\rightarrow 7S_{1/2}$ transition
in cesium, for which
$E_1^{pv} \simeq 10^{-11} i\; ea_{0}$ and $\vert M_1 \vert \simeq 2 \times 10^4\;\vert E_1^{pv} \vert$.

\section{APV in the highly forbidden cesium transition: Past results}

\subsection{General principle of the Cs experiments} 
In order to observe the atomic transition well above the background coming from loosely bound cesium dimers, we apply a static
electric field $\bf E$. This field induces a transition electric dipole ${\bf d}^{ind}$ via
parity-conserving mixing of atomic P states with S states. In this induced dipole
\begin{equation}
{\bf d}^{ind} = \alpha {\bf  E }+ i \;\beta \;{\vec \sigma} \times {\bf E}\;,
\end{equation}
there are actually two contributions, one due to the scalar transition polarizability $\alpha$, and
the other to the vector polarizability $\beta$. Here ${\vec \sigma}$ stands for the  electron
spin operator. The coupling of ${\bf d}^{ind}$ to the electric radiation field $\hat \epsilon_{ex}$ 
provided by a resonant laser wave, gives rise to an induced electric dipole amplitude, $E_1^{ind}=
{\bf d}^{ind} \cdot \hat \epsilon_{ex}$. An excellent control of $E_1^{ind}$ can be achieved by
adjusting the strength of ${\bf  E }$, its direction with respect to both the direction of the light
beam and the beam polarization $\hat \epsilon_{ex}$. Reversing the direction of {\bf E} can even
give a specific signature to all effects linear in $E_1^{ind}$. For all these reasons $E_1^{ind}
E_1^{pv}$ has become the kind of interference effect detected in all cesium APV experiments so
far. The transition rate $ \vert E_1^{ind}\vert ^2$ grows like the square of the
electric field, while the asymmetry $E_1^{pv}/E_1^{ind}$ is inversely proportional to the field, but
still can reach values above $10^{-6}$.

Experiments only measure amplitude ratios like $Im\;E_1^{pv}/\beta E$ and since $\beta$ is
known from atomic theory only to within 1$\%$ accuracy, there was a crucial need for an
accurately known amplitude usable for {\it absolute calibration}.
Fortunately, as shown in 1988 by our ENS-Paris group\cite{bou881,bou882}, a very precisely
known transition amplitude does exist. In zero electric field the $M_1$
$6S_{1/2}
\rightarrow 7S_{1/2}$ Cs transition amplitude contains a certain contribution $M_1^{hf}
\simeq 0.19
\times M_1$, induced by the off-diagonal hyperfine interaction. Experimentally this
contribution is easily identified because it contributes only to the $\vert \Delta F \vert = \pm 1$
hyperfine lines and with opposite signs. 
The key-point is that, up to corrections less than 0.25\% coming from many-body effects\cite{bou881,joh99,dzu00},
$M_1^{hf}$ can be expressed in terms of the geometrical mean of the diagonal matrix elements\cite{bou881}, {\it i.e.} the hyperfine splittings $\Delta W$ of the two S states, themselves
measured in cesium with high precision: $M_{1}^{hf}=C \sqrt {\Delta W_{6S} \cdot \Delta W_{7S}}$.
Finally, as advocated in ref.\cite{bou882}, a precise measurement of $M_1^{hf}/\beta$ then provides
an {\it absolute} determination of $\beta$ thereby making possible an absolute determination of
$E_1^{pv}$ with a theoretical uncertainty less than $2.5 \times 10^{-3}$. 

As a last step,  
Atomic Physics calculations are essential to extract $Q_W$ from the experiment.
The quantity
$ { E}^{pv}_1 $ can be considered as an infinite sum over the intermediate
$ P $
states admixed with the $ S $ states by the parity-violating
interaction\nobreak\ :
$$ { E}^{pv}_1=   \sum^{ }_ n{
\left\langle 7S_{1/2} \right\vert d_z \left\vert nP_{1/2} \right\rangle
\left\langle nP_{1/2} \right\vert V_{pv} \left\vert 6S_{1/2}
\right\rangle \over E \left(6S_{1/2} \right)-E \left(nP_{1/2} \right) }+\ {\rm
crossed\ terms} \; . $$
The atomic orbitals and the valence-state energies are perturbed by
many-body
effects. Initially, two
different approaches have been employed, one semi-empirical\cite{bou864},  and the second starting from
first-principles\cite{dzu891,blu90}, following techniques inspired from field theory to perform relativistic many-body computations (see $\S 4.1$ for more details). Results agree within the stated precision.

\subsection{Completed Cs experiments: the very first in Paris, and a high precision one in Boulder} 

The experiment completed in Paris\cite{bou82}, in 1982 and 1983, was the first carried
out in cesium. The detected signal was the {\it circularly polarized} fluorescence on the 
$7S_{1/2}-6P_{1/2}$ transition following circularly polarized 6S-7S excitation in a transverse {\bf  E } field. 
With 12 $\%$ experimental accuracy and a theoretical uncertainty at that time
less than 8 $\%$, it has led to a quantitative test of the standard electroweak theory in the
electron-hadron sector, which is of a new kind\cite{bou97}. Because of the low momentum
transfer involved ($\simeq$ 1 MeV/c)  it extends considerably the range of $\vert q^2 \vert$
where the theory finds experimental support. Moreover, information deduced from $Q_W$
complements that obtained from high energy experiments because, as mentioned above (Eq. 3), in
atoms all the nucleons and all the quarks act coherently\cite{bou83}. 
In particular, it provides a test for additional neutral
vector bosons. This point will be discussed for the present context in the last section. 

APV can also be a source of valuable information relevant for nuclear
physics, via the nuclear-spin-dependent PV interaction. The relevant
experimental parameter involved here, is the difference between the two asymmetries
measured on two different hyperfine lines belonging to the same transition\footnote {\small The
cesium natural isotope having the nuclear spin 7/2, the $6S_{1/2}$ and $7S_{1/2}$ states
possess two hyperfine substates with total angular momentum $F=$3 and 4.}.
If non-zero, the quantity:
\begin{equation}
r_{hf}=\frac {A_{LR} (6S \rightarrow 7S, \Delta F=-1)} {A_{LR} (6S \rightarrow 7S,
\Delta F=+1)} -1 \; ,
\label{Rhf}
\end{equation} is a manifestation of  the nuclear-spin-dependent interaction.
An effect of about $4\%$ is expected as a result of the parity violating interactions taking place
inside the nucleus between the quarks. The valence electron is contaminated by photon exchange (see the
corresponding diagram represented on Fig. 1, right). The theoretical concept relevant for describing this
effect is the nuclear anapole moment, introduced a long time ago by Zel'dovich\cite{zel57}. For a
simple interpretation in terms of chiral magnetization of the nucleus the reader is referred to\cite{bou99}. Explicit calculations have been performed for cesium by the Novosibirsk\cite{fla80} and  Paris\cite{bou911} groups. It is interesting to note that the effect
of  the electron-nucleus
$Z_0$ exchange associated with an axial-vector coupling to the nucleons, also involving the
nuclear spin, is formally identical but about five times smaller.

In 1997 precision measurements of $Im\;E_1^{pv}/\beta$ on the two hyperfine lines $\Delta F = \pm1$ were made by the Boulder group\cite{woo97}. 
 This experiment operates on
a spin polarized atomic beam that crosses an optical power build-up cavity resonant for the 6S-7S transition.
DC electric and magnetic fields, ${\bf E}$ and 
${\bf B}$ are applied perpendicularly to the angular momentum, $\xi
{\bf k}$, of the resonant light beam. The pseudo-scalar quantity which manifests PV
is the mixed product: $ {\bf E} \times {\bf B} \cdot \xi {\bf k}$. It is expected to
appear in the transition rate of single Zeeman components of the transition.
One monitors a change in the ground state population induced during passage through the interaction region which is correlated with the parameter reversals. 
Combining the results quoted for the two hyperfine lines $\Delta F = \pm 1$, 
one can extract the
nuclear-spin-independent contribution: 
$$Im \; E_1^{pv}/\beta=(-1.5963 \pm 0.0056 \rm) \; mV/cm,$$
discussed later on, as well as the $r_{hf}$ parameter (Eq. \ref{Rhf}): 
$$r_{hf}=(4.8 \pm 0.7) \times 10^{-2}\;.$$
This last result provides a clear manifestation of the nuclear anapole moment. However, it appears inconsistent (roughly by a factor of 2) with other data relating to PV nuclear forces\cite{bou911,fla04}. So an independent measurement looks necessary.

Using their very precise measurement of $M_1^{hf}/\beta$ performed more recently\cite{ben99} (quoted accuracy of 0.12 $\%$) and the atomic theory at that time, 
the result for the weak charge finally reported by the Boulder group is :
$$Q_W^{exp}=-72.06 \pm 0.28_{exp} \pm 0.34_{th}\;. $$
Note that the theoretical uncertainty of 0.4\% quoted here is 2.5 times smaller than the 1\% uncertainty estimated by the theoreticians, but it was assigned by the group based on the ability of the theoretical models to reproduce atomic test parameters\cite{ben99}. When the errors of different origins are added in quadrature,
the fractional accuracy is 0.6$\%$.
This result has to be compared to the SM prediction\cite{mar84}, recently updated\cite{roe}:
$$Q_W^{th}=-73.19 \pm 0.13.$$
A deviation between experiment and theory of $1.13 = 2.6\;\sigma$ was concluded.
The existence of an extra neutral gauge boson\cite{lan99}, whose mass lies in the range hundreds of GeV, was invoked as  
a possible interpretation in terms of "new physics" beyond the Standard Model, that did not contradict high energy physics results.

\section{New activity in the field}
\subsection{Atomic theory}

The 0.4$\%$ theoretical uncertainty assigned on $Q_W^{exp}$ by the Bouder group raised questions
about small corrections to the prediction neglected so far. This prompted several theoretical
groups to reconsider the problem.

A. Derevianko\cite{der01} was the first to evaluate the Breit
correction (the magnetic interaction  between all electrons) and to announce a non-negligible
correction to $Q_W^{exp}$ $\simeq -0.6\%$,  rapidly confirmed by other groups\cite{koz00,fla02}.
The effect of a difference between the neutron and proton distributions, already examined\cite{blu90}, 
was revisited and confirmed to be small for cesium,
$-0.2\%$ with an uncertainty $\lesssim 0.1\%$\protect\cite{fla02}$^{-}$\protect\cite{sil}.

Theorists of relativistic many-body calculations have refined their calculations and still agree on the result within a 1$\%$ level of precision\cite{koz00,koz02,fla02}, with a precision of 0.5$\%$\cite{fla02} being claimed. In addition, the last     calculations now include the contribution of self-energy and vertex QED radiative corrections\cite{kuc02,mil01} found to be non-negligible (-0.85$\%$).

Once reinterpreted at the light of the latest theoretical results, 
$Q_W^{exp}$ becomes:
$$Q_W^{exp}=-72.71 \pm 0.29_{exp} \pm 0.39_{th}\; .$$ The current deviation
$Q_W^{exp}-Q_W^{th}$ is thus less than one
$\sigma$\cite{fla04}. In view of this new result, the lower limits
on the mass of a possible $Z'$ boson have been reanalyzed\cite{cas02} and found comparable to
those deduced from the four LEP experiments.

In spite of this apparently perfect agreement with the SM, we would like to underline two
reasons (one theoretical and the other experimental) why it may be somewhat too early to consider
such an agreement as definitely well established.

\begin{romanlist}[(ii)]

\item A slight risk of double counting in the radiative-correction
evaluation: 
not only the atomic factor but also $Q_W^{th}$ include QED radiative corrections. 
A global calculation incorporating all corrections to the
parity violating electron-nucleus interaction would seem to be more rigorous.

\item Necessity of independent measurements.
\begin{itemlist}
\item Cross-check of the empirical ratio $\beta/M_1^{hf}$ seems important.
When the precise Boulder determination of $\beta/M_1^{hf}$ is used to obtain a value for the vector polarizability $\beta$, the result differs from a recent semi-empirical determination of
$\beta$\cite{vas02} by $(0.7 \pm 0.4) \%$. Though small, such a
difference, considered alone, is sufficient to shift the value of $Q_W^{exp}$ by $1.3\;\sigma$\cite{fla02}.

\item An independent determination of $Q_W^{exp}$ using a totally different method would be welcome as a cross-check.
Our current experiment developed at ENS-Paris should actually fulfill this objective.
 \end{itemlist}

\end{romanlist}

\subsection{A novel experimental approach: APV measured in Cs using
 stimulated-emission detection and asymmetry amplification}
Our new experiment currently performed on the Cs $6S \rightarrow 7S$ transition differs radically from both our first 82-83 experiment\cite{bou82} and the Boulder experiment\cite{woo97}.
While all APV experiments on highly forbidden transitions in a Stark field have
used so far the detection of fluorescence signals, this experiment exploits the possibilities of the
$\Lambda$-type three-level system $6S_{1/2}-7S_{1/2}-6P_{3/2}$ in interaction with two copropagating lasers.
One intense laser with linear polarization $\hat \epsilon_{ex}$ excites the forbidden transition in
a longitudinal {\bf E} field and the second laser probes the angular anisotropy induced in the
excited state. The probe laser is amplified and its linear polarization $\hat \epsilon_{pr}$ is altered.
A precise analysis of the polarization change is performed on the transmitted beam. In order to
provide suitable conditions for PV measurements, a large gain for the probe has to be achieved,
hence the realization of this experiment with a pulsed excitation laser and a gated probe. The
left-right asymmetry is a rotation of the linear polarization of the probe $\hat \epsilon_{pr}$,
which can be measured at each laser pulse with a dual-channel polarimeter operating in balanced
mode (see Fig. 2).
The effect is the manifestation of the pseudoscalar $(\hat
\epsilon_{ex} \cdot \hat \epsilon_{pr})( \hat \epsilon_{ex} \wedge \hat \epsilon_{pr}\cdot {\bf
E})$ present in the optical gain, i.e. a {\it chiral} term which is responsible for contributions of opposite signs to the amplified intensity detected in both channels\cite{gue03}.

This {\it chiral} optical gain can be understood on the basis of simple symmetry considerations. The
excitation polarization $\hat \epsilon_{ex}$ and the {\bf E} field determine two symmetry planes
for the experiment. Without parity violation one would expect the excited Cs vapor to have its
optical axes contained in those planes. If this were the case, a probe beam linearly polarized with
$\hat \epsilon_{pr}  \parallel \hat \epsilon_{ex}$ would pass through the vapor without
alteration of its polarization. Actually, due to parity violation acting during the excitation process,
the optical axes of the excited vapor are tilted with respect to the symmetry planes and it is this
tiny tilt angle
$\theta^{pv} = -Im \; E_1^{pv}/\beta E \simeq 10^{-6}$ rad, odd under {\bf E}-reversal, which has to
be determined. As a consequence of this tilt, while the probe beam passes through the
vapor its  polarization rotates towards the axis of larger gain. This causes an imbalance at the
output of the polarimeter, odd in {\bf E}-reversal, which is measured. It is precisely calibrated by
measuring the imbalance induced by a small precisely known tilt of $\hat \epsilon_{ex}$, in identical conditions.
\begin{figure}[t]
\centerline{\psfig{file=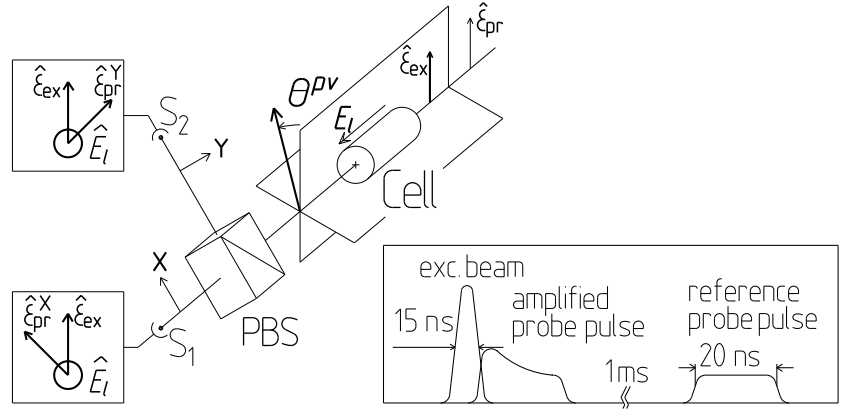,width=5.0in} }
\vspace*{-2cm}
\caption{Schematic of the ENS-Paris pump-probe Cs experiment showing the two orthogonal symmetry planes defined
by the electric field ${\bf E}$ and
the linear excitation polarization  $\hat
\epsilon_{ex}$. APV gives rise to a tilt
$\theta^{pv}$  of the optical axes of the excited vapor out of those planes. The incoming probe polarization 
$\hat \epsilon_{pr}$ provides a
superposition of the right- and left-handed ($\hat \epsilon_{ex}, \hat \epsilon_{pr}, {\bf E}$)
configurations analyzed. The probe amplification difference is
directly extracted from the optical signals S$_1$, S$_2$, recorded in each channel of the Polarizing Beam Splitter
(PBS). Inset: timing of the experiment repeated at $\sim\;150\;Hz$. 
(Fig. adapted from Gu\'ena {\it et al.}\protect \cite{gue03}).}
\end{figure}

An attractive feature of this experiment is that {\it the left-right asymmetry} itself, $A_{LR}$, is
amplified while the probe propagates through the optically thick vapor. Therefore, instead of
being a decreasing function of the applied field as in usual fluorescence experiments, $A_{LR}$ is
transformed by stimulated-emission detection into an increasing function of ${\vert \bf E \vert}$\cite{bou96}.
Another conceptual
difference with the Boulder experiment is that the detected observable is directly the asymmetry itself as opposed to a
modulation of order $6
\times 10^{-6}$ in the total transition rate, correlated with parameter reversals. Moreover, the present approach avoids the difficulty met there
regarding systematics associated with $M_1$ interference effects  and line-shape dependent effects. Another attractive feature is the cylindrical symmetry 
exhibited by the experiment: the signal is expected to remain invariant under simultaneous rotations of the laser beam polarizations $\hat \epsilon_{ex}, \hat \epsilon_{pr}$ about the common beam direction. This feature enables us to discriminate against possible systematics arising from stray transverse fields and misalignments\cite{bou04}. 

However, the observation of the PV {\it chiral} optical gain remains an experimental challenge\cite{gue98}. For instance, detection has to be restricted to the $\sim 20$ ns time interval during which the pulse excited vapor acts as an amplifier. This is achieved using a fast optical switch on the probe laser beam, with severe requirement with respect to its extinction ratio\cite{gue05b}. Among several other unusual problems to be solved, there was the generation of a pulsed longitudinal uniform {\bf E} field of 2kV/cm inside a 8-cm long Cs vapor cell at the useful atomic density and submitted to intense laser light pulses. One key-element here was provided by the all-sapphire so-called ''grooved'' cells, finally upgraded with high optical quality metal coated windows. An original trick to enhance the signal has been the use of a polarization-tilt magnifier.

\begin{figure}[t]
\centerline{\psfig{file=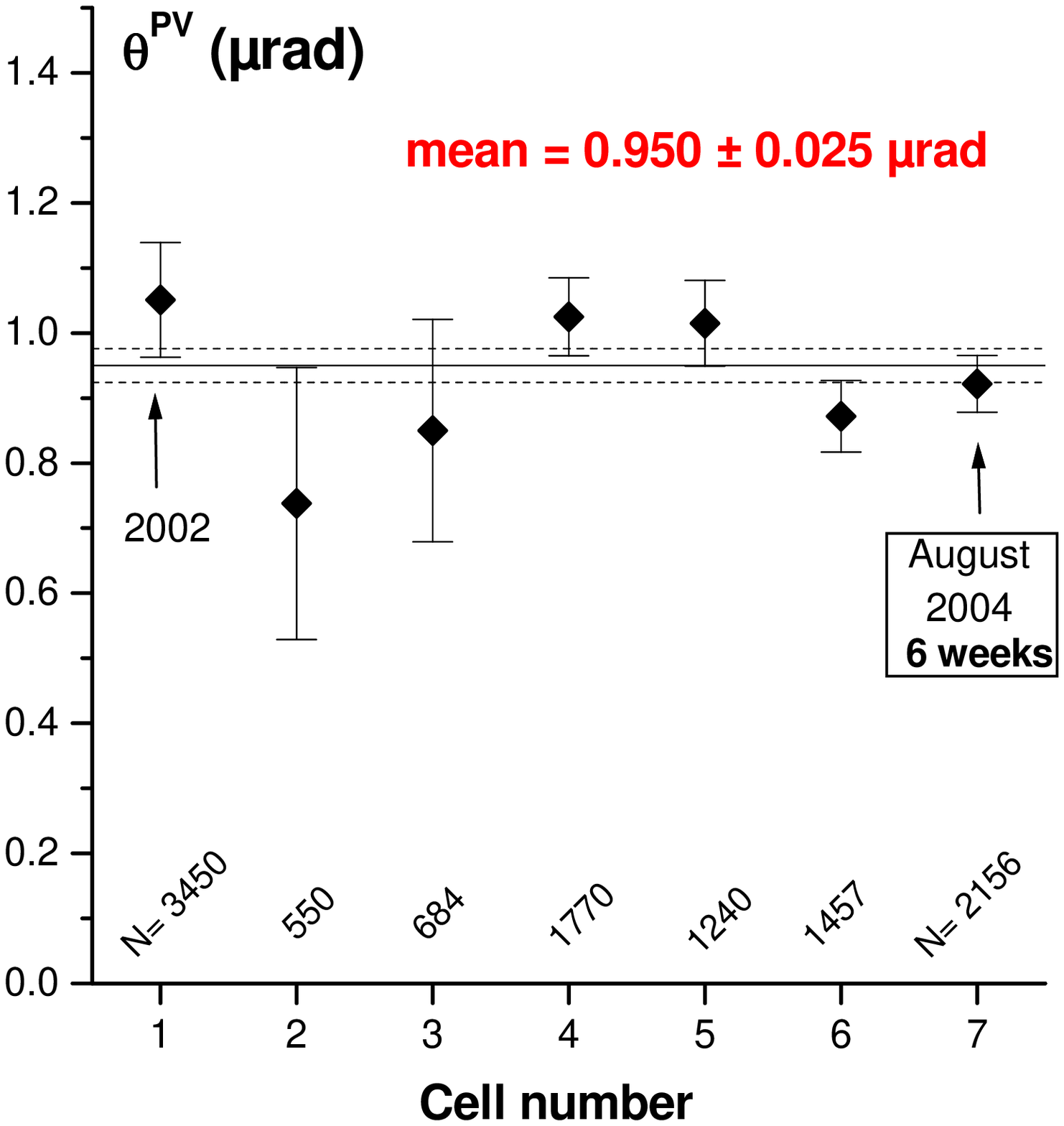,width=2.0in} \hfil \psfig{file=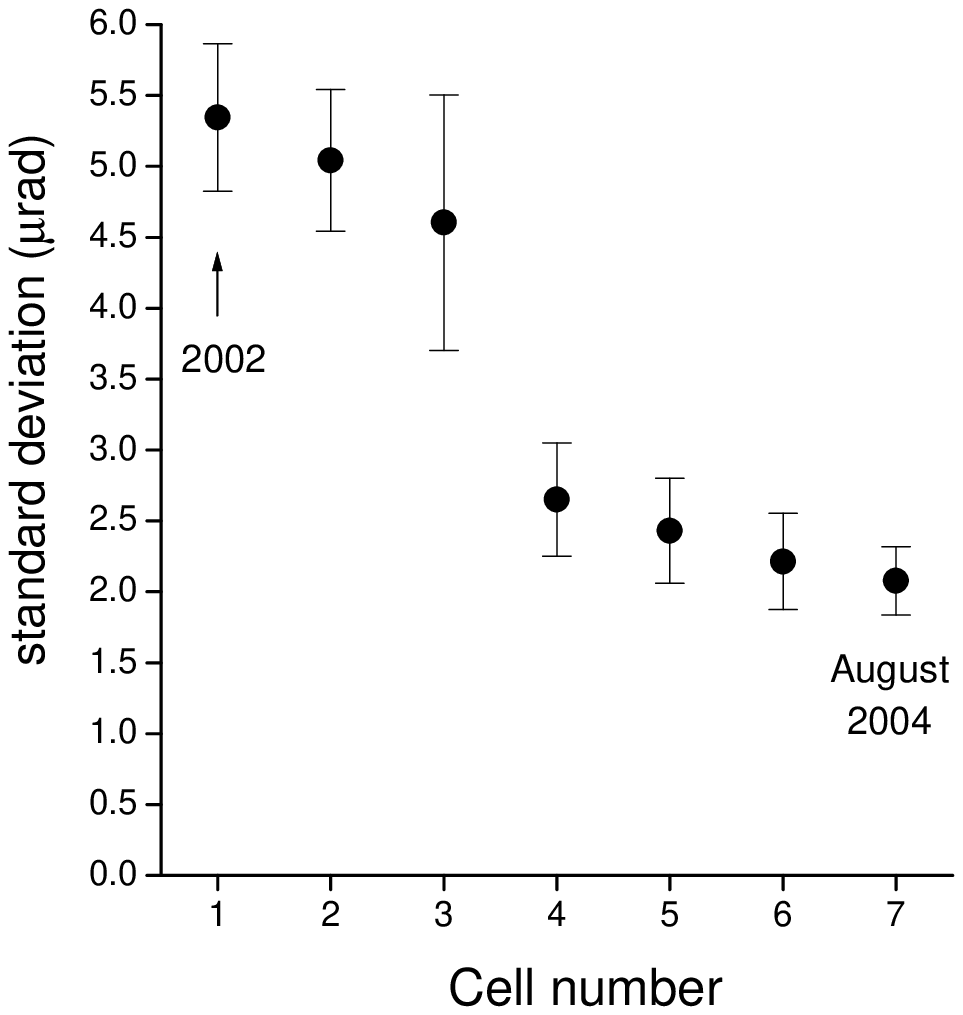,width=2.0in}}
\vspace*{4pt}
\caption{Results of the ENS experiment. Left: Values of $\theta^{pv} = -Im \; E_1^{pv}/\beta E$ obtained in different Cs cells: mean (diamond) and statistical error (error bar) from the sample of N data acquired in each of the 7 cells. Solid and dashed lines: global mean and statistical error, respectively. Right: Standard deviation of the distribution of the PV data relative to each cell. Adapted from Ref.\protect \cite{gue05b}.}
\end{figure}
Fig. 3, Left summarizes the experimental determinations of $\theta^{pv} = -Im \; E_1^{pv}/\beta E$ obtained in seven different cesium vapor cells, which are well compatible ($\chi^{2}$=7.7 for 6 degrees of freedom). 
The experiment provides for real-time tests of the systematic effects and consistency tests in the data reinforcing confidence in the results\cite{gue05b}. Using our 1\% accurate {\it in-situ} determination\cite{gue05b} of the Stark induced amplitude $\beta E$ in terms of the precisely known (see $\S\;3.1$) $M_1$ amplitude, 
our present result for the $6S_{F=3}\rightarrow 7S_{F=4}$ hyperfine component is
$$Im \; E_1^{pv}=(-0.808 \pm 0.021)\times 10^{-11}\; ea_{0}$$
or, relying on the determination of $\beta$ in Ref.\cite{ben99}:
$$Im \; E_1^{pv}/\beta=(-1.538 \pm 0.040 \rm)\; mV/cm\;.$$
It is affected with negligible systematic uncertainty and reaches a 2.6$\%$ statistical accuracy. It agrees well with the  Boulder result $Im \; E_1^{pv}/\beta=(-1.558 \pm 0.008 \rm)\; mV/cm$ for the same component. The {\it absolute} precision of $2\times 10^{-13}ea_{0}$ for $Im \; E_1^{pv}$ 
is already 10 times better than that achieved in other heavier atoms (Tl, Pb and Bi)\cite{bou97}. 
 
Fig. 3, Right exhibits the improvement on Signal-to-Noise from the first\cite{gue03} to the last cell : it implies that the averaging time  needed to reach a given statistical accuracy is reduced by a factor of $\sim 12$. 
This gain in sensitivity has been obtained by our exploiting the process of asymmetry amplification by stimulated emission.
The level of sensitivity is now adequate for a one-to-two percent $Q_W$-precision objective. There is still room for improvement by boosting the amplification phenomenon,
for instance in a new \textit{transverse} field geometry as we have proposed recently\cite{gue05}. In this proposal, a 0.1\% statistical precision looks achievable.

The work reported in this section has demonstrated an original method for parity violation measurements in a highly forbidden atomic transition offering perspectives for high precision and reliability.

\subsection{ Experiments in progress in other atoms and new proposals }
\subsubsection{Work in progress on a chain of rare earth isotopes }
It has not been possible, yet, to test experimentally an important prediction of the SM concerning the
variation of $Q_W$ along a string of isotopes. It has been suggested\cite{dzu86} that in rare
earth spectra one can find atomic states of opposite parity which are nearly degenerate, for instance in Dy (Z=66) and Yb (Z=70) which both 
offer a chain of seven stable isotopes. One expects this near-degeneracy to enhance the PV effect, thus making possible precise
measurements. In first approximation, ratios of $E_1^{pv}$ amplitudes  should provide
ratios of the weak charges, without invoking atomic physics calculations\cite{for90}, made complex by configuration mixings.

  The search in Yb, conducted in Berkeley\cite{ytt},
presents analogies with that performed in Cs. The spectrum of this atom is much simpler than that of other rare earth atoms, so that the theoretical predictions should be
more reliable. In the chosen transition, 
$E_1^{pv}$ is predicted to be 100 times larger than in Cs. Important
exploratory work has already been carried out and the $M_1$ amplitude of this highly forbidden transition has already been measured\cite{bud02}.

\subsubsection{Prospects with cooled and trapped atoms}
Cooling and trapping techniques open the way to measurements with radioactive cesium
isotopes or even francium atoms. With Z=87, Fr is expected to lead to PV effects 18 times larger than Cs ($Z^3$ law and relativistic effects\cite{bou97}, $\S\;3.1$).
In addition, Fr has many isotopes.

Fr atoms, either obtained from a radioactive source or produced on-line by an accelerated ion
beam colliding a target, are produced at a limited rate with a superthermal
velocity distribution. The first prerequisite is to avoid their spreading out in space and loss
inside the wall. Successive attempts to load Fr atoms in a neutral atom trap have already made
possible the observation of several Fr allowed transitions, leading to precise
spectroscopic measurements\cite{wie97,oro00}. In our opinion, the observation of the forbidden
$6S \rightarrow 7S$ line with a sample of cold Cs atoms would represent an important
preliminary step to assess the feasibility of a PV measurement in the Fr $7S \rightarrow 8S$ forbidden transition (for a definite proposal, see\cite{san03}).

\subsubsection{Static manifestations of the electroweak interaction}
When an atom is placed in a chiral environment, Sandars' theorem\cite{san} no longer holds
and Parity Violation can manifest itself by an energy shift of its atomic levels in spite of T-invariance. Although shifts of this
kind still require considerable experimental efforts to be detected, their existence presents undoubted conceptual interest.

For chiral molecules one expects a small energy shift
between the two mirror-image enantiomers. It has been
searched for by comparing vibrational frequencies of the right and left-handed species of
CHFClBr molecules\cite{bor}. Experiments in more favorable conditions are in progress.

When Cs atoms are trapped in a solid matrix of $^4$He of hexagonal symmetry, two applied
{\bf E} and {\bf B} static fields and the crystal axis $\hat n$ create a chiral
environment around each atom. In these conditions, a linear Stark shift proportional to the
Cs nuclear anapole moment and to the (T-even) P-odd pseudoscalar
$(\hat n \cdot {\bf B}) (\hat n \cdot {\bf B} \wedge {\bf E})/B^2$ has been predicted\cite{bou01}.

Both effects could provide {\it a static manifestation of the electroweak interaction}, which is still
missing.

\section{Relevance of Atomic Parity Violation today}


The main goal of Atomic Parity Violation\cite{bou97} is to provide
a determination of the weak nuclear charge $Q_W$, from the measurement of
$E_1^{pv}$ {\it via} an atomic physics  calculation which now aims at
0.1$\%$ precision in cesium\cite{der04}. In view of present
 and forthcoming results from high energy experiments, an important issue concerns the relevance of further
improving difficult experiments such as APV measurements. We would like to present arguments in favor of
small scale APV experiments.

\begin{romanlist}[(ii)]
 \item
 First  we wish to reiterate  that APV experiments explore the
 electroweak (EW) electron-hadron  interaction  within a range of low momentum transfers $q_{at}$
 of 1 MeV or there abouts in cesium, as opposed to the huge ones explored in collider experiments:
 100 GeV at LEP~I and LEP~II and 1 TeV at LHC. 
To obtain relevant information, one has to approach
an absolute precision of $10^{-8}$ in the measurement of a radiative atomic transition left-right asymmetry.
 \item
 For $q_{at} \sim 1$ MeV, the quarks of the atomic nucleus act
 {\it coherently}, while at high energies the nucleons are broken into their
fundamental constituants:
the quarks act then {\it incoherently}. This is what happens in deep inelastic
electron-nucleon  scattering, such as the SLAC experiment\cite{pre78} involving a
 GeV polarized electron beam colliding against a  fixed deuterium target. As a consequence,
different  combinations of  electron-quark  PV  coupling constants are involved in the LR asymmetries
of  the two experiments:
  $\frac{2}{3}C_u^{(1)} -\frac{1}{3}C_d^{(1)}$ at high energies instead
 of  $( 2Z + N) C_u^{(1)} + (Z+ 2N) C_d^{(1)} $ for $Q_W$.
 It is easily seen that, in a model-independent analysis, the two experiments delimit nearly orthogonal allowed bands in the
$\lbrack  C_u^{(1)},C_d^{(1)} \rbrack $ plane\cite{bou97}.
\item
 The fact that $q_{at}\sim $1 MeV allows one to investigate the
 possible existence of a \textit{light} extra neutral gauge boson,
  for instance with a mass in the range  of a few MeV. Such a drastic
modification of EW interactions  appears as  an alternative explanation
for the remarkably intense  and narrow   gamma  ray line emitted from the bulge of our galaxy, close to the energy  of 511
keV which coincides with the electron mass\cite{sil04,boe04}. According to this
somewhat exotic  model, the observed spectrum
 would result from the annihilation of two \textit{light dark matter particles}
 (mass $\geq$ 1-2 MeV) into a pair $(e^{+}, e^{-})$ via the
 exchange  of a light gauge boson U, with a mass
 of about 10  MeV\cite{fay04}.
 In order to reproduce the size of the observed effect,
 one  has to  exclude at a large confidence level an
 {\it axial} coupling of the electrons to  the new  U boson, while such a coupling
is  the only possible   one for \textit{dark particles} which carry no charge.
This is  where APV comes into play.

The most plausible conclusion to which the present value\cite{fla04} of $\Delta Q_W$ leads is that
the U boson couples  to the electron  as a vector particle with no axial coupling  at the $10^{-6}$ level,
while its vector coupling to leptons and quarks are of the same order of magnitude\cite{bou04b}. Thus, the APV
measurements  provide  an empirical justification
 for  a key-hypothesis, introduced  in  the astrophysical model accounting for the $ 511$ keV galactic line.
 \item
 Deviations $\Delta Q_W$ of $Q_W^{exp}$ from the SM prediction, are
 most often analyzed in the framework of  {\it "new physics"} models
 which affect EW interactions {\it at energies higher than $M_{Z_0}
 c^2$} through the existence of gauge bosons heavier than the $Z_0$, such as for
 instance Kaluza-Klein excitations of the SM gauge bosons\cite{ant01}.  It turns out that
 $\Delta Q_W$ is proportional to the same factor
 $X= \frac{\pi^2}{3 n} R_{\parallel}^2 M_{Z_0}^2$ as the deviations from
the SM  in  existing collider experiments, provided that $q^2 R_{\parallel}^2 \ll 1$, where
 $R_{\parallel} \leq 1$ TeV$^{-1}$  stands for the compactification
 radius associated with the additional $d_{\parallel}$ dimensions
 of the new physical space for EW gauge fields.
A determination of
 $\Delta Q_W$  below the  0.1 ${\%} $ level of precision would give
 constraints on $R_{\parallel}$, competitive with those of LEP II\cite{delg00,all04,bou04a}.
Furthermore, one can consider models which predict effects undetectable by LEP II results but that would be
visible in APV experiments\cite{delg00,che03}. Therefore, a 0.1$\%$ accurate determination of $Q_W^{exp}$ could allow
one  to impose a $\sim 5$ TeV limit to the compactification mass $R_{\parallel}^{-1}$  in a direction possibly invisible to high energy experiments.
\end{romanlist}

In view of the present need for further measurements, underlined above, there are strong incentives
to pursue APV measurements in Cs by stimulated-emission detection and to push further the technologies opening the path to novel-type experiments, both in cold radioactive atoms and chains of rare earth isotopes.
We find it remarkable that results of APV experiments that involve scattering photons, of only
a few eV, by a sample of a few cubic centimeters of dilute atomic vapor, can stand comparison with experiments
performed in colliders of the  highest energy, for providing a lower limit on the mass of a hypothetical additional neutral boson.

\section*{Acknowledgments}
Partial support from IN2P3 (CNRS) and from BNM-INM is gratefully acknowledged. Laboratoire Kastler Brossel is a Unit\'{e} Mixte de Recherche de l'Universit\'{e} Pierre et Marie Curie et de l'Ecole Normale Sup\'erieure, associ\'{e}e au CNRS (UMR 8552). F\'{e}d\'{e}ration de Recherche
de l'ENS is associ\'{e}e au CNRS (FR684).


\end{document}